\documentclass[DIV=12]{scrartcl}
\usepackage{appendix}
\usepackage{graphicx}
\usepackage{float}
\usepackage{textcomp,gensymb}
\usepackage{amsfonts}
\usepackage{amsmath}
\usepackage{mathtools}
\usepackage{amssymb} 
\usepackage{adjustbox}
\usepackage{multirow}
\usepackage{array}
\usepackage{siunitx}
\usepackage{lmodern}
\usepackage[style=phys,sorting=none,maxbibnames=6,minbibnames=3,url=false,date=year,giveninits,doi=false,url=false]{biblatex}
\usepackage{xcolor}
\usepackage{caption}
\usepackage{subcaption}
\usepackage{wrapfig}
\usepackage[inline]{enumitem}
\usepackage{csquotes}
\usepackage{authblk}

\title{LET-modifying joint optimization for mixed-modality photon-proton treatment planning}

\author[1,2]{Lisa Seckler}
\author[1,3]{Amit Ben Antony Bennan}
\author[1,3]{Niklas Wahl}
\affil[1]{Division of Medical Physics in Radiation Oncology, German Cancer Research Center (DKFZ), Heidelberg 69120, Germany}
\affil[2]{University of Applied Sciences (THM), Giessen, 35390, Germany}
\affil[3]{Heidelberg Institute for Radiation Oncology (HIRO), Heidelberg 69120, Germany}

\begin{document}

\maketitle

\section*{Abstract}
As depth increases, linear energy transfer (LET) rises toward the distal edge of the Bragg peak, boosting the radiobiological effectiveness (RBE).
To manage the biological variation and limit normal-tissue damage, LET-modifying objective functions on, e.\,g., dose-weighted LET or \enquote{dirty dose} and/or usage of variable RBE models were introduced. Because shaping LET by proton irradiation alone has its limits, this work proposes to jointly optimize mixed-modality proton-photon treatments based on directly LET-modifying objective functions.  

The investigated objective functions rely on either dose-weighted LET or dirty dose concepts. To formulate a consistent combined optimization problem, the contribution of secondary electron LET in photon treatments is considered (and discussed) as well. Combined dose/LET calculation and optimization are realized in the open source toolkit matRad. Phantom plans as well as a patient plans are optimized for analysis on the method, combining five proton fractions with 25 photon fractions. Thereby, dose-optimized combined plans are used as a reference.

The reference plan shows that protons are, in general, dosimetrically superior and thus preferred, with photons aiding in achieving conformity. The introduction of LET modified objectives locally modifies the proton contribution in the targeted regions of interest. Especially at the distal edge, the photon contribution increases to move high-LET / dirty dose out of the OARs. Dirty dose objectives seem to allow a more comprehensive steering of the high-LET regions compared to LETxDose.

Ultimately, incorporating LET-based objectives into a jointly optimized proton-photon system allows for improved dose conformity and reduced high-LET exposure in critical regions in proximity to the distal proton edge. This approach enables the utilization of modality-specific strengths and can contribute to safer, more effective treatment plans.

\section{Introduction}
In a proton beam, the linear energy transfer (LET) increases with depth and is highest at the distal edge of the Bragg peak. Although clinically a constant Radiobiological Effectiveness (RBE) of 1.1 is used to scale the absorbed physical dose to a biological (photon equivalent) dose, in reality, an increase in LET leads to a higher RBE \cite{Rorvik2018,mcnamara_phenomenological_2015,paganetti_report_2019}. Accounting for this increased differential biological effectiveness is desirable in targeting the tumor site to achieve local control, on the other hand, the higher effectiveness is also undesirable when talking about dose to target-adjacent organs at risk (OAR)\cite{oden_spatial_2020,henjum_organ_2021}. Furthermore, there is no consensus on the biological effectiveness of protons at the end-of-range and hence uncertainty in the predicted RBE within OAR \cite{giovannini_variable_2016}.  

Thus, recent work has been investigating the direct use of LET in plan optimization to act as a surrogate for RBE \cite{giantsoudi_linear_2013,deng_critical_2021,mcintyre_systematic_2023}. \textcite{Grun2019} show that the dose-averaged LET is a sufficient predictor of RBE (for regions of a narrow LET spectrum). 
The introduction of LET-based objectives along with RBE-weighted plan optimization was also suggested as a method to reduce interpatient biological variability for proton treatment plans \cite{McMahon2018,liu_robust_2020,cao_linear_2017,unkelbach_reoptimization_2016,giantsoudi_linear_2013,giovannini_variable_2016}. Another proposed approach applies an objectives to control the placement of proton track-ends in critical structures \cite{traneus_introducing_2019}.
Within the target volume, LET-based treatment planning has been suggested with the concept of LET painting of radio-resistant hypoxic tumors to increase tumor control probabilities \cite{Bassler2014,Kopp2020}.

\begin{figure}[h!]
    \centering
    \includegraphics[width=0.7 \textwidth]{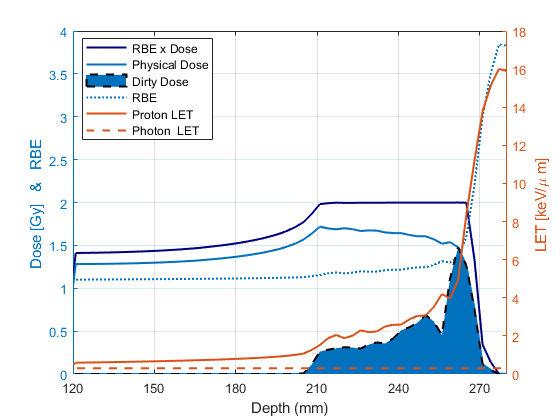}
    \caption{Dirty-Dose component (LET threshold = $2 \ keV/\mu m$) of physical dose in a Spread out Bragg peak optimized with the McNamara variable RBE model. Also shown on the axis on the right, dose averaged LET of Protons and the constant LET assumed for photon beams .}
    \label{fig:SOBP}
\end{figure}

Recently, in an attempt to quantify this high distal LET combined with dose, the concept of dirty-dose was introduced  \cite{heuchel_dirty_2024}. This idea assumes that dose leads to critical biological damage in combination with a high LET value. Dirty-dose describes the dose delivered by particles of an LET greater than a predetermined threshold, as visualized in Figure \ref{fig:SOBP}. Although there is no consensus on the definition of a high LET threshold, this approach may be a pragmatic means to assess and address the effects of high LET radiation. 

The introduction of LET-based objectives, either direct or via indirect quantities like dirty dose, introduces new tradeoffs in the multi-criteria optimization problem balancing target coverage and OAR sparing inherent to treatment planning. In the case of LET limiting objectives for OAR, usually the trade-off is seen in target dose coverage. In the case of LET objectives for the Target volume, the trade-off can be visualized as the increase in integral dose from the entrance channel and target dose conformity \cite{oden_spatial_2020}.

Attaining a uniform or high LET in the target volume is typically a challenge for large target volumes. Multi ion treatments, where combination of ion therapy beams are used in a single fraction, have been investigated by \textcite{Kopp2020,kramer_overview_2014} in order to optimize for higher LET distributions within the target volume. Here the combination of different radiation qualities allows for greater flexibility in shaping the LET distribution within the target volume.

Here we hypothesize that a mixed modality approach combining photons with proton relying on LET modifying objectives may circumvent the presented limitations of proton only LET-based treatment planning. The combined mixed modality approach presented in here jointly optimizes the beamlet (photons) and spot (proton therapy) intensities simultaneously in order to achieve the cumulative prescribed dose where each fraction is delivered using a single modality \cite{Unkelbach2021,Bennan2021,Fabiano2020,Fabiano2020a,Unkelbach2018a,torelli_new_2024,Marc2021}. This gives the optimizer the freedom to generate heterogeneous fraction dose distributions using the available radiation modalities. Naturally these additional degrees of freedom would offer the opportunity to modulate and control the increased LET and RBE at the end of range of particle therapy treatments by preferentially using lower LET dose, possibly from photons to achieve conformal plans.

Using simultaneously optimized mixed-modality treatments (photon-proton treatments and photon-carbon ion treatments), this work aims to 
\begin{enumerate*}[label=(\arabic*)]
    \item compare the approaches of LET weighted dose optimization, dirty-dose optimization and variable RBE based optimization to mitigate heightened end-of-range effects in target-adjacent OAR, and
    \item demonstrate mixed modality treatments to optimize for a homogeneous LET weighted dose and dirty-dose in the target volume 
\end{enumerate*}

\section{Methods and Materials}

\subsection{Joint optimization problem}
Jointly optimized plans are optimized on a type of cumulative dose. To consider different fractionation properties, often a biological dose like the biological effective dose (BED)\cite{Unkelbach2018a} or equivalently the biological effect $\varepsilon = -\ln(S) = BED\times\alpha$ \cite{Bennan2021} given the survival fraction $S = e^{-\alpha d - \beta d^2}$ from the linear-quadratic (LQ) model. This allows for consistent biological accumulation of the possible heterogeneous fraction dose distributions within the scope of the LQ model. The corresponding optimization problem can be stated as follows:

\begin{equation}
    \label{eq:EffectJointOptimization}
    \begin{aligned}
    & \underset{w}{\text{minimize}}
    & & \mathcal{F} (w) = \sum_m p_m \; f_m (\boldsymbol{\varepsilon}^{\textrm{Total}}(w)) & \\
    & \text{subject to} & & \varepsilon^{\textrm{Total}} _i = n^\gamma \varepsilon_i^\gamma + n^P\varepsilon^P_i & \forall i\\
    & & & \varepsilon_i^\gamma = \alpha_i^\gamma \sum_j D^\gamma_{ij} \; w^\gamma_j + \beta_i^\gamma \left( \sum_j D^\gamma _{ij} \; w^\gamma _j \right) ^2 & \forall i \\
    & & & \varepsilon^P_i = \sum_k \alpha^P _{ik}  D^P _{ik} \;  w^P_k + \left( \sum_k \sqrt{\beta^P _{ik}} 
     D^P _{ik} \; w^P_{k} \right) ^2 & \forall i \\ 
    & & & w^\gamma_j \geq 0  & \forall j\\
    & & & w^P_k \geq 0 & \forall k\\
    \end{aligned}
\end{equation}

The objective function $\mathcal{F}(\boldsymbol{\varepsilon}^{Total})$ is defined as the sum of individual objectives $f_m(\boldsymbol{\varepsilon}^{Total})$, each based on the total biological effect $\boldsymbol{\varepsilon}^{Total}$, and is weighted by a penalty factor $p_m$. 
$\varepsilon_i^{Total}$ in voxel $i$ is the sum of the fraction biological effects from photons $\varepsilon_i^{\gamma}$ and particles $\varepsilon_i^{P}$ multiplied by their respective number of predetermined fractions, $n^\gamma$ and $n^P$. 
The intensity of an individual photon beamlets $j$ is represented by $w_j^\gamma$, whose dose contribution to voxel $i$ per unit intensity is stored in the dose influence matrix element $D_{ij}^\gamma$. Similarly, $w_k^P$ represents the intensity of the particle therapy beamlet $k$, and $D_{ik}^P$ denotes its physical dose contribution to voxel $i$ per unit intensity. The parameters $\alpha_i^\gamma$ and $\beta_i^\gamma$ are the reference LQ model coefficients for photons. The effective LQ model parameters for particle therapy, $\alpha^P_{ik}$ and $\beta^P_{ik}$, depend on the spatial location of voxel $i$ along the path of beam $k$. 

For a constant RBE assumption for proton therapy, the effective $\alpha$ and $\beta$ are assumed to be equal to the photon LQ model parameters while the dose is multiplied with a factor of \num{1.1}. The objective function is optimized to determine the set of pencil beam intensities for both photons ($w_j^\gamma$) and particle therapy ($w_k^P$) that produces the optimal distribution of the total biological effect.

In order to maintain the readability of the plans and reduce the number of variables for the analysis in this work, we assume no fractionation benefit ($\alpha/\beta_{Healthy} \geq \alpha/\beta_{Tumor}$), i.\,e., photon LQ model parameters $\alpha = 0.1$ and $\beta = 0.05$ for both tumor tissue and healthy tissue. 

The number of fractions allocated to each modality is not part of the optimization problem but pre-decided based on boost treatments \cite{Combs2010a}.

This work utilizes objective functions ($f_m$) that are based on total biological effect and LET based quantities, namely, dirty dose and LET weighted dose.

\subsection{Introducing LET-modifying objectives}
  When interacting with a medium, energy loss straggling broadens the energy spectrum, and nuclear reactions generate new secondary particles. To fully characterize the beam at a specific location, it is essential to determine the particle energy spectrum.
  Here we utilize the dose averaged LET ($LETd$), as it is widely used for the prediction of RBE in proton therapy\cite{cao_linear_2017,Rorvik2018}.

\subsubsection{LET-weighted dose (LETxDose) objectives}
LETxDose is LET weighted by the absorbed dose and is a measure of the additional biological dose from high LET \cite{unkelbach_reoptimization_2016}. For a more computationally efficient implementation, a single LET weighted dose contribution matrix($L_{ij}$) can be created such that

\begin{equation}
    L_{ij} = LET_{ij} \cdot D_{ij}
    \label{eq:LETxDose}
\end{equation} 
Here $L_{ij}$ is the LET weighted dose contribution of beamlet $j$  in voxel $i$.

For the LET distribution, however, it must be noted that the LET is a characteristic of each type of radiation and can not be offset against the number of fractions. Therefore, for the total LET value, we approximate it by averaging over dose across the entire treatment. This treatment spanning  LET distribution can be given as a "quasi" $LET_d$: 

\begin{equation} 
    LETd_i^{\textrm{quasi}} = \frac{n^\gamma L_{ij}^\gamma \cdot w^\gamma_j  + n^P L_{ij}^P w_j^P  }{n^\gamma D_{ij}^\gamma \cdot w^\gamma_j+  n^P D_{ij}^P w_j^P}
    \label{eq:Gesamt-LET}
\end{equation}

where additional the index $P$ denotes the particle therapy modality component, while the index $\gamma$ defines the photon component. The variable $n$ stands for the number of fractions of the respective beam type.

\subsubsection{Dirty Dose (DD) objectives}
Dirty dose is the component of the physical dose delivered by high LET ($LET_d$) areas of a proton beamlet. The dirty-dose concept from \textcite{heuchel_dirty_2024}, the dirty-dose used here is only dependent on an LET-threshold. The dirty-dose influence matrix($D^{\textrm{Dirty}}$) can be written mathematically as:
\begin{equation}
    D^{\textrm{Dirty}}_{ij} = \begin{dcases}
        D_{ij} & \text{for} \ LET_{ij} >  k^{Dirty}\\
        0      & \text{else} 
    \end{dcases}
\label{dij.DD}
\end{equation}
in which $D^{\textrm{Dirty}}_{ij}$ describes the dirty-dose contribution for each voxel $i$ from pencil beam $j$. $D^{\textrm{Dirty}}_{ij}$ is obtained by applying a threshold to the LET contribution matrix $LET_{ij}$ and sampling the dose contributions $D_{ij}$  for voxel - pencil beam $(i,j)$ pairs with LET contributions greater than the LET threshold $k^{Dirty}$ .

\subsection{Treatment planning}
Initial testing and sensitivity analysis for the LET objectives were investigated for the TG119 phantom. 
The objective settings for the sensitivity analysis are shown in table \ref{Objectives1}.
\begin{table} [H]
 \caption[Basic dose objectives for all plans.]{Basic dose objectives of the structures of the TG119 phantom that remain for each treatment plan.}
\begin{center}
\begin{tabular}{ c| c| c }
\hline
 structure & primary objective & secondary objective\\
 \hline
 \hline
 Target & SD(800,60) \\ 
 Body & SO(100,30) & MD(100,0) \\  
 OAR & SO(100,30)\\ 
\hline
\end{tabular}
\label{Objectives1}
\end{center}
\end{table}
The dirty dose as well as the LETxDose objectives are applied for the OAR. The objectives of the modified LET objectives are indicated in the results figures \ref{fig:dose_DD_LxD} and \ref{fig:comp_LETxDose}.

The developed methods were then demonstrated on a clinically relevant case, an Anaplastic astrocytoma patient case from the Glioma image Segmentation dataset (GLIS-RT) \cite{shusharina_cross-modality_2021}. This patient case was chosen due to the location of the target volume adjacent to the brain stem. Additionally, the GTV is considered as a boost volume for an integrated boost treatment.  

We considered 5 fractions of intensity modulated proton treatment (IMPT) ($265\degree$, $285\degree$), and 25 fraction of photon (intensity modulated radiotherapy (IMRT)) (nine equidistant beams) for the treatment plan. 

Dose objectives were created to generate a basic treatment plan, as seen in figure \ref{table:BasicObjectives}.  A squared deviation objective (SD) is applied on the target, whereas squared overdosing (SO) is used for the OARs \cite{Wieser2017,Wieser2018a}. A mean dose objective (MD) is also added as a secondary objective in the body in order to minimize the integral dose.  This reference plan contains standard objectives that remain in every plan (Table \ref{table:BasicObjectives}).
\begin{table} [H]
 \caption[Standard objectives]{Primary treatment planning objectives (same for all plans).}
\begin{center}
\begin{tabular}{ c| c| c| c }
\hline
 structure & objectives & penalty & prescribed dose\\
 \hline
 \hline
 CTV & SD & 800 & 60 Gy\\ 
 GTV & SD & 1000 & 68 Gy \\
 Brain stem & SO & 100 & 50 Gy\\ 
 Chiasm & SO & 200 & 30 Gy \\
 Body & SO & 100 & 30 Gy \\
 ~ & MD & 100 & 0 Gy\\
 
\hline
\end{tabular}
\label{table:BasicObjectives}
\end{center}
\end{table}

The plans described in this study:
\begin{enumerate}
    \item Proton only plan using a constant RBE model (Only proton). This plan is intended to show the current standard of single modality treatment.
    \item Simple jointly optimized plan that use the McNamara variable RBE model for protons \cite{mcnamara_phenomenological_2015} (Reference Joint). This plan is intended to showcase the degrees of freedom afforded by jointly optimized plans. 
    \item Jointly optimized plan that uses dirty-dose objectives (Joint + DD )
    \item Joint optimized plan that uses LET weighted dose (LETxDose)  objectives (Joint + LET)
\end{enumerate} 

For the purpose of optimization, look-up tables of LETd are used from the machine models available with matRad that were generated from Monte Carlo simulations for all available energies of protons. The LET values are calculated for each voxel and pencil beam pair and stored in a LET contribution matrix $LET_{ij}$. This model assumes a simplistic constant lateral LET for steps in depth. The LET contribution matrix $LET_{ij}$ is used for all LMO objectives\cite{cao_linear_2017}.

LET modifying objectives are added to the primary planning objectives depending on the type of plan. Squared overdosing objectives were used in the OARs as seen in \ref{tbl:OverdosObj}  and squared underdosing objectives were used in the GTV (table \ref{tbl:UnderdosObj}). 
\begin{multline}
    f^{\textrm{LMO sq overdosage}} = \frac{1}{N_S} \cdot \sum_{i\in S} \Theta(q^{ref}(LET_i,d_i) - q_i(LET_i,d_i)) \cdot \\ (q^{ref}(LET_i,d_i) - q_i(LET_i,d_i))^2
    \label{LMOSqO}
\end{multline}

\begin{multline}
    f^{\textrm{LMO sq underdosage}} = \frac{1}{N_S} \cdot \sum_{i\in S} \Theta( q^{ref}(LET_i,d_i) - q_i(LET_i,d_i)) \cdot \\ (q^{ref}(LET_i,d_i) - q_i(LET_i,d_i))^2
    \label{LMOSqU}
\end{multline}
where $q_{i}(LET_i,d_i)$ is the LET modifying quantity in voxel $i$. For dirty-dose, $q_{i}(LET_i,d_i) = D_{ij}^{\textrm{Dirty}} w_j$ and for LET weighted dose $q_{i}(LET_i,d_i) = L_{ij} w_j$.

For dirty-dose computations, we utilize a LET threshold value of $2 \ keV/\mu m$  within the range suggested by \textcite{heuchel_dirty_2024} ($1 - 4 keV/\mu m$) for protons.

For LET weighted dose computations a constant LET value of $0.3 \ keV/\mu m$ of the secondary electrons is attributed to the photons for the purpose of optimization \cite{Murshed.2019}. 

\begin{table} [H]
    \caption[LMO objectives]{LET modifying objectives that are added to the reference plan  in the brain stem for the squared Overdosing LMO.}
   \begin{center}
   \begin{tabular}{l| c| c| c| c }
   \hline
    Plan type & OAR & Objectives & Penalty & Prescribed dose\\
    \hline
    \hline
    Ref Joint & Brain stem & Mean Dose & 100 & 0 Gy\\ 
    DD Joint & Brain stem & DD SO & 100 & 0 Gy\\
    LET Joint & Brain stem & LETxD SO & 100 & 20 Gy\\ 
   \hline
   \end{tabular}
   \label{tbl:OverdosObj}
   \end{center}
   \end{table}

\begin{table} [H]
    \caption[LMO objectives]{LET modifying objectives that are added to the reference plan  in the GTV for the squared Underdosing LMO.}
   \begin{center}
   \begin{tabular}{l| c| c| c| c }
   \hline
    Plan type & OAR & Objectives & Penalty & Prescribed dose\\
    \hline
    \hline
    DD Joint & GTV & DD SU & 100 & 18 Gy\\
    LET Joint & OAR & LETxD SO & 100 & 90 Gy\\ 
   \hline
   \end{tabular}
   \label{tbl:UnderdosObj}
   \end{center}
   \end{table}

The open source treatment planning platform \emph{matRad}(v2.10.1) \cite{Wieser2017,Wieser2018a,Ackermann2020} was used to implement joint optimization of combined treatments and LET-based optimization objectives \cite{Bennan2021,Bennan2022}.

\section{Results}
\subsection{LET modifying objectives in the OAR}

\subsubsection{Sensitivity of dirty-dose objectives}
\begin{figure}[H]
    \centering
    \includegraphics[width=0.8 \textwidth]{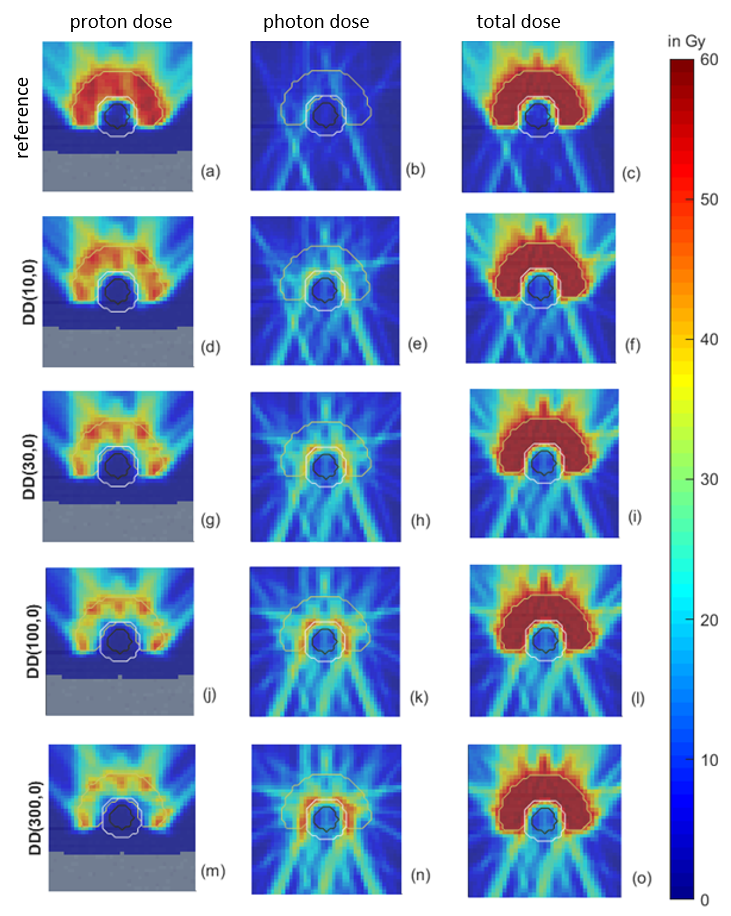}
    \caption{Comparison of proton and photon contribution to the total dose considering dirty-dose objectives with regard to the reference plan.}
    \label{fig:dose_DD_LxD}
\end{figure}

To compare the proton and photon contribution of a reference plan and a LET modifying plan, it is necessary to look at the total dose and each individual radiation modality. In figure \ref{fig:dose_DD_LxD} you can see the reference plan in the first row and additional plans with dirty-dose objectives added to the core and a bigger core, named core-big, in the lines below. 
To compare proton and photon dose contributions, the dose per fraction is multiplied by the number of fractions for each modality. 

\subsubsection{Sensitivity of LETxDose objectives}
\begin{figure}[H]
    \centering
    \includegraphics[width=0.8 \textwidth]{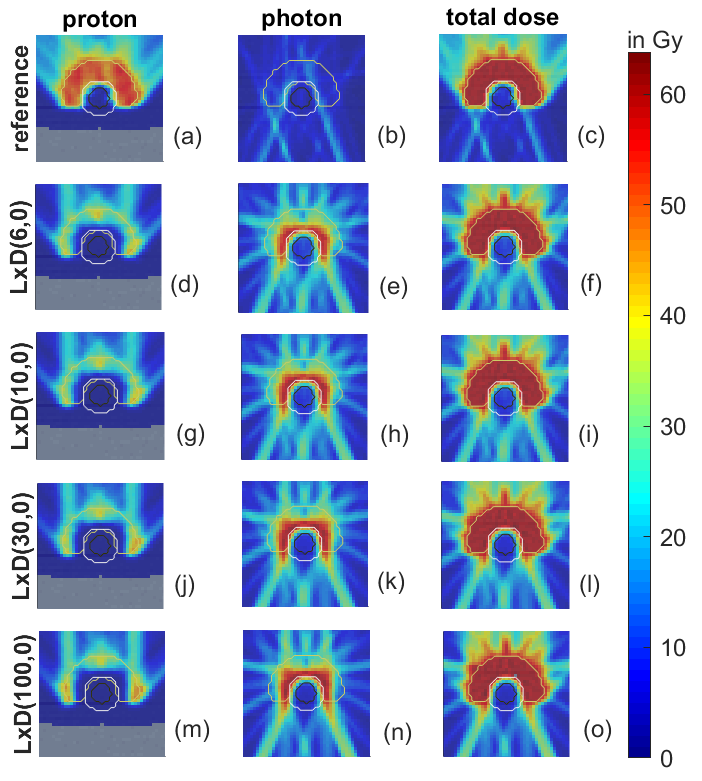}
    \caption{Comparison of proton and photon contribution to the total dose considering LETxDose objectives with regard to the reference plan.}
    \label{fig:comp_LETxDose}  
\end{figure}

A comparison is also made for additional plans with LETxDose objectives, seen in figure \ref{fig:comp_LETxDose}. 

\subsubsection{Reference plans}
The only proton plan delivers a uniform 2~Gy in the CTV and 2.26~Gy in the GTV with an integrated boost and, naturally, delivers lowest mean dose to surrounding tissue due to the lack of photons.

In contrast, the reference jointly optimized plan delivers most of the dose within the 5 proton fractions and uses photon fractions to improve dose conformity and reduce the dose in the proton entrance channel. The improvement in dose conformity from the additional degrees of freedom of the jointly optimized plan can be seen in the reduction of the near-maximum dose in the brain stem ($BED_{2\%}$) in Table \ref{tbl:OverDoseQI}.

\begin{figure}[h!]
    \centering
    \includegraphics[width=1\textwidth]{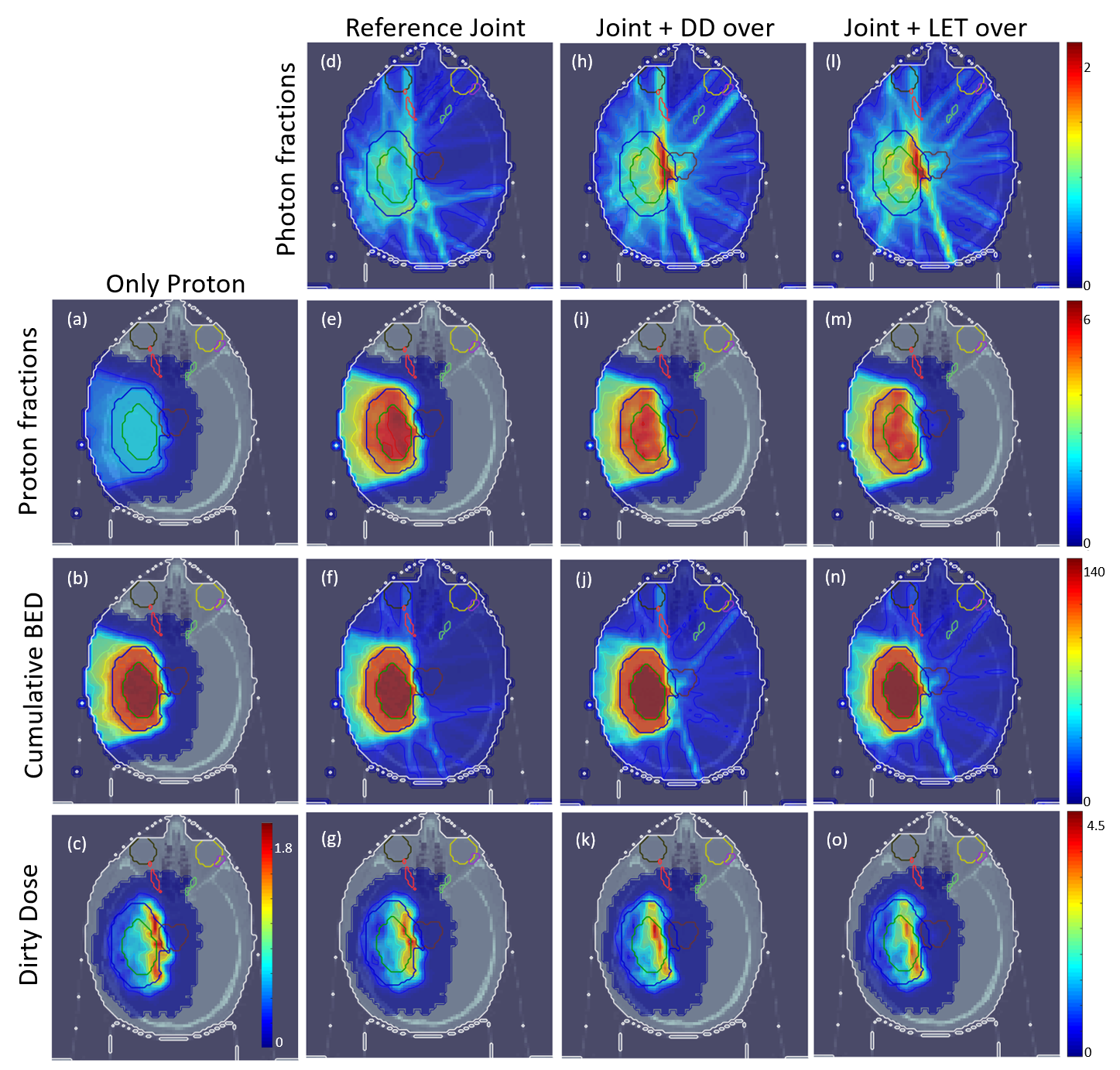}
    \hfill
    \caption{Visualized photon fractions (d, h, l), proton fractions (a, e, i, m), cumulative BED (b, f, j, n) and dirty-dose (c, g, k, o) for an only proton plan (a, b, c), a reference jointly optimized plan (d, e, f, g), a joint plan with a DD overdosing objective (h, i, j, k) and a jointly optimized plan with a LET overdosing objective (l, m, n, o). Dirty dose shown here is also dependent on the fraction dose, hence the only proton plan (2 Gy per fraction) has an adjusted color scale for dirty dose seen in subfigure (c) }
    \label{fig:allSlice}
\end{figure}

\begin{figure}[h!]
    \centering
    \includegraphics[width=1\textwidth]{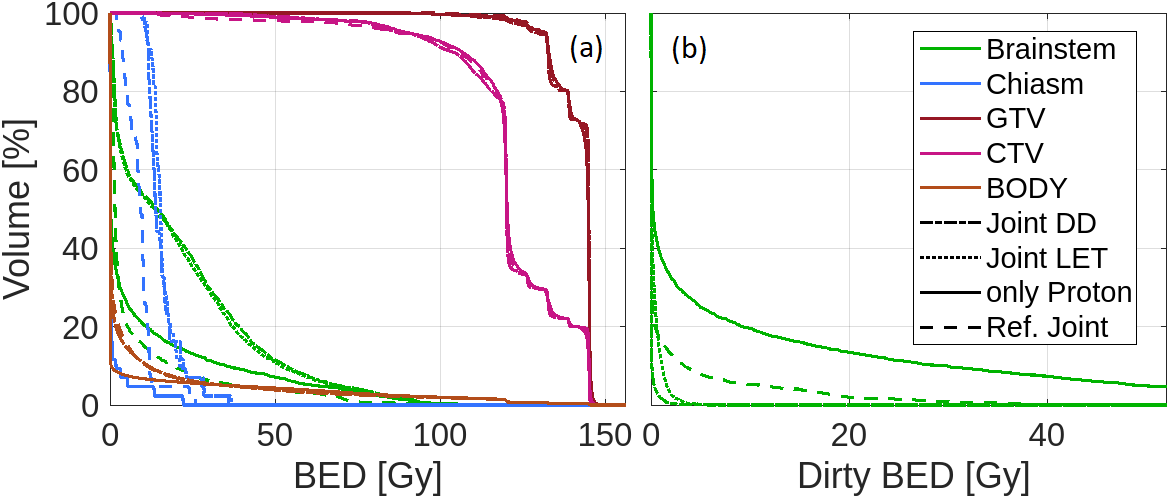}
    \hfill
    \caption{Modified dose volume histograms for plans with LET Modifying objectives for OAR a) BED Volume histogram b) Dirty Dose volume histogram for the brainstem.  }
    \label{fig:VHS}   
\end{figure}

\subsubsection{LET-modified joint plans}
Figure \ref{fig:allSlice} compares the joint treatment plans obtained with LET-modifying objectives on the OARs to the reference plans (only proton, reference joint) (see Table \ref{tbl:OverdosObj}). All plans were comparable in target dose coverage as shown in the cumulative BED-volume histograms in Figure \ref{fig:VHS}.
For the jointly optimized plans with LMO's, the interface between the CTV and the Brainstem receives increased photons dose. The interface region receives comparable doses from both photons and proton fractions (approximately 2 Gy per fraction). The cumulative $BED_{2\%}$ in the Brainstem for both plans is comparable to the only proton plan. However, due to the increased use of photons, LMO utilizing plans seem to trade LET modification with a higher mean dose in the Brainstem. 

$D^{\textrm{Dirty}}_{2\%}$ (Dirty Dose to 2 \% of brain stem volume cumulated over 30 fractions) for the only proton plan is far greater than that from Joint + DD over  and Joint + LETxDose  (both over 5 fractions). On the fraction scale, the only proton plan delivers 2~Gy per fraction and the LMO Joint optimized plans deliver, approximately, 5~Gy to the target and one would assume that the dirty dose delivered as only proton plan would deliver a less dirty dose. However, the fraction dirty dose from only proton plan is greater than that from DD Joint plan and LETxDose Joint plan by a factor of 7.5 and 3.3, respectively. This already suggests a benefit in using LET modifying objectives. The cumulative BED and dirty BED  plan quality metrics over the entire treatment can be found in Table \ref{tbl:OverDoseQI} 

\begin{table}[H]
    \begin{tabular}{l|c|c|c|c}
        \hline
    Cumulative BED  & Only  & Reference  & Jointly Opt.  & Jointly Opt\\
      & Proton & Jointly Opt. & Dirty Dose & LET x Dose  \\
        \hline
        \hline
    GTV $BED_{95\%}$    & 131.3    & 131.9      & 129.6     & 129.7 \\
    CTV $BED_{95\%}$    & 96.3      & 97.6      & 94.6      & 94.4  \\
    Brain stem $BED_{2\%}$      & 83.9   & 69.4  & 87.8      & 85.8  \\
    Brain stem Mean    & 9.5         & 6.8       & 21.3      & 20.6 \\
    Body Mean         & 2.7         & 3.8       & 3.6       & 3.5 \\
    \hline
    Cumulative Dirty BED & & & & \\
    \hline
    Brain stem Mean          & 7.6   & 1.6   & 0.06  & 0.35 \\
    Brain stem $BED_{2\%}$   & 62.4  & 19.9  & 0.9   & 2.3 \\
    Brain stem std. dev.     & 15.9  & 5.16  & 0.3   & 0.7 \\
    \hline         
    \end{tabular}
    \caption{Table of treatment plan quality indicators based on cumulative BED and cumulative dirty-dose for plans with LMO applied to OAR }
    \label{tbl:OverDoseQI}
    \end{table}

\subsection{LET modifying objectives in the Target}
\begin{figure}[!h]
    \centering
    \includegraphics[width=0.7\textwidth]{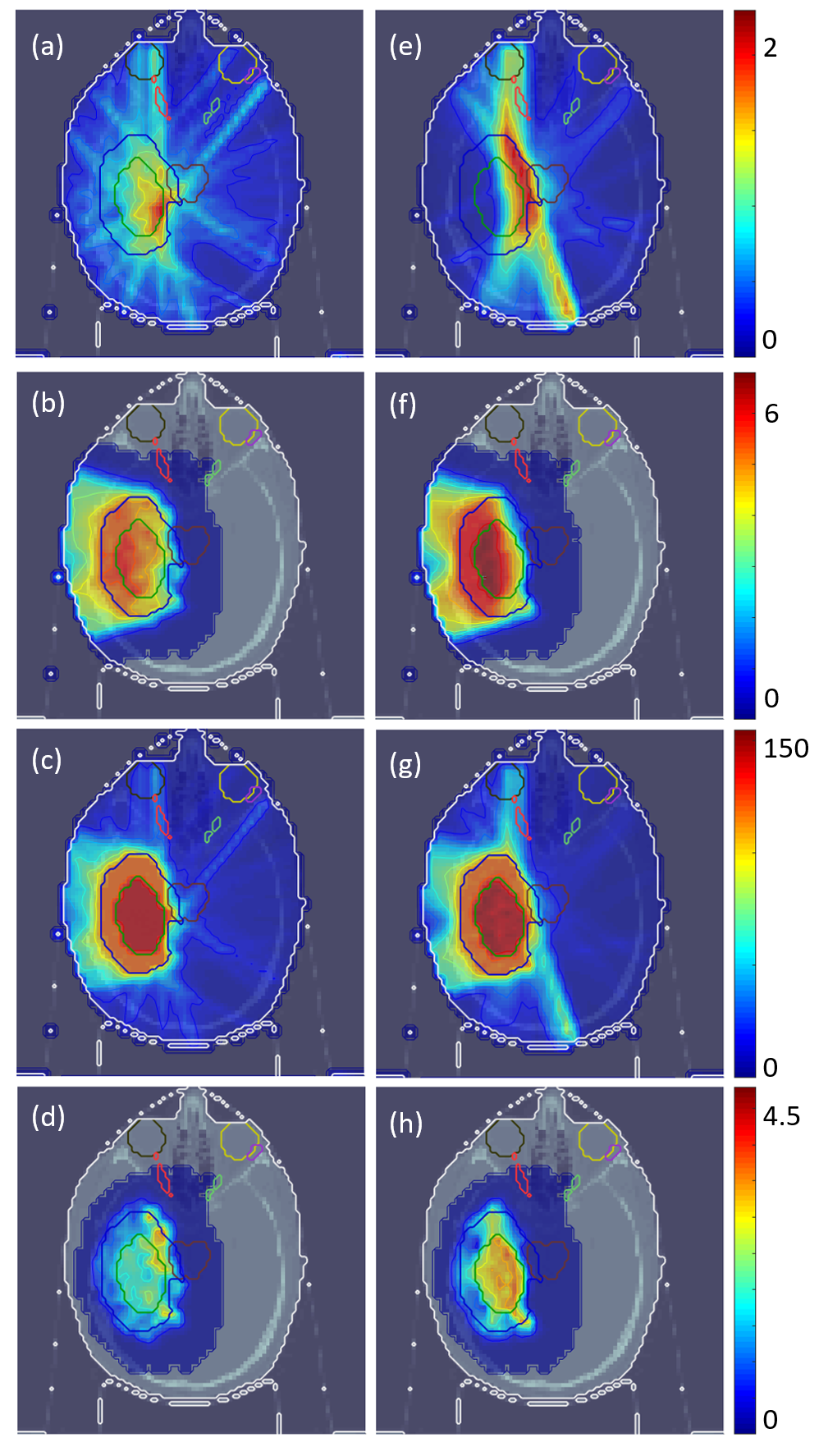}
    \hfill
    \caption{Visualized photon fractions (a,e), proton fractions (b,f), cumulative BED (c,g) and dirty-dose (d,h) for an a jointly optimized plan with a DD underdosing objective (a,b,c,d) and a jointly optimized plan with a LETxDose underdosing objective (e,f,g,h). }
    \label{fig:allSlice_Under}
\end{figure}

To demonstrate the impact of LMO in the Target, LETxDose and dirty-dose objectives were formulated as squared underdosing objectives for the GTV (Table \ref{tbl:UnderdosObj}), in addition to the primary treatment objectives(Table \ref{table:BasicObjectives}). As before, the plans are compared against a 30 fractions only proton treatment and a simple jointly optimized proton-photon treatment (Figure \ref{fig:allSlice}). The plans with LMO for the GTV can be seen in Figure \ref{fig:allSlice_Under} . Here, proton dose is concentrated at the proximal side of the GTV. The gradual dose fall-off within the GTV results high LET within the GTV. However, this also leads to dose cold spots in the Target. Photon dose is used to make up for dose regions which are not irradiated using protons.

Dosimetric quality indicator for the compared plans can be seen in Table \ref{tbl:UnderdoseQI}. Although the only proton plan shows a higher cumulative near-maximum dose within the GTV, the LMO affected plans show a lower variation in the high LET dose within the GTV. Furthermore, the plan with Joint + dirty-dose underdosing delivers a higher mean dirty-dose. On the other hand, the Joint + LETxDose plan shows the lowest delivered dirty-dose values as the LETXDose objective does not distinctly control high LET component of the dose. 

\begin{table}[H]
    \begin{tabular}{l|c|c|c|c}
        \hline
        Cumulative BED               & Only  & Reference  & Jointly Opt.  & Jointly Opt\\
                    & Proton & Jointly Opt. & Dirty Dose & LET x Dose  \\
        \hline
        \hline
    GTV $D_{95\%}$        & 131.3    & 132.0   & 131.9  & 132.2  \\
    PTV $D_{95\%}$        & 96.3     & 97.6    & 102.0  & 98.5 \\
    Brain stem $D_{2\%}$   & 83.9     & 69.4    & 94.1   & 93.6 \\
    Brain stem mean        & 9.5      & 6.8     & 21.1   & 31.5 \\
    Body mean             & 2.7      & 3.8     & 4.2    & 5.2  \\

     \hline
       Cumulative Dirty BED & & & & \\
    \hline
    GTV mean       & 32.4    & 20.2    & 40.9     & 18.6 \\
    GTV $D_{2\%}$  & 64.7    & 47.9    & 52.4     & 27.9 \\
    GTV std.dev    & 12.4    & 11.1    & 7.7      & 3.8 \\

    \hline

    \end{tabular}
    \caption{Table of treatment plan quality indicators based on cumulative BED and cumulative dirty-dose for plans with LMO applied to the Target }
    \label{tbl:UnderdoseQI}
    \end{table}

\begin{figure}[!ht]
    \centering
    \includegraphics[width=1\textwidth]{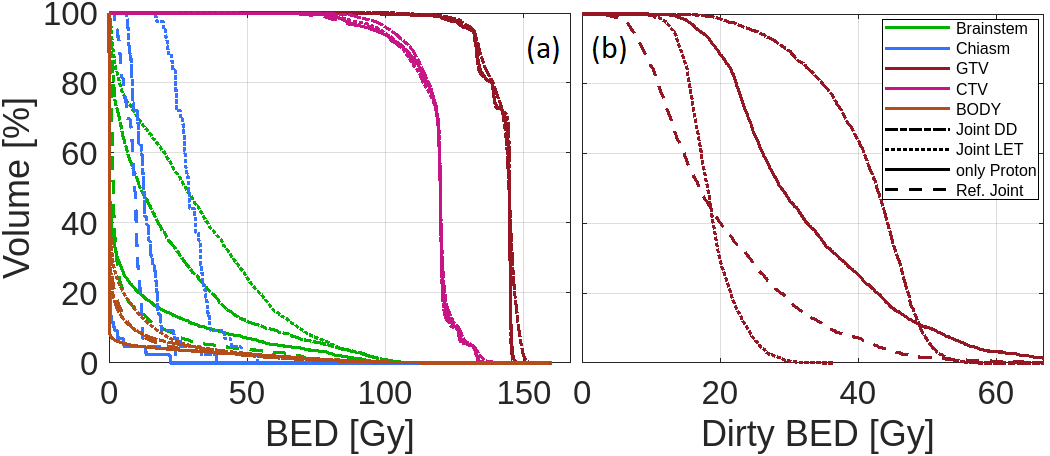}
    \hfill
    \caption{Modified dose volume histograms for plans with LET Modifying objectives for the Target a) BED Volume histogram b) Dirty Dose volume histogram for the GTV. }
    \label{fig:Target_VHS}   
\end{figure}

\newpage
\section{Discussion}
This study explores the implementation of LET-based objectives for OARs and the Target, within the jointly optimized combined treatment framework. In order to do this, two reference plans were considered: (1) only proton plan with integrated boost, which would present the case for current state of the art proton therapy treatments (2) jointly optimized photon-proton plan, that would present the case for an unencumbered jointly optimized mixed modality treatment.
In general, with regards to a jointly optimized plan, the mean dose objective in the BODY drives the optimization towards a proton plan to minimize the integral dose. Photons were used to reduce the dose in the entrance channel and to improve the dose conformity with the additional degrees of freedom.

The assumption ($\alpha/\beta_{Target} <= \alpha/\beta_{Normal Tissue}$) of no biological fractionation benefit was made in order to maintain the explainability of the plans by avoiding the use of photons for the purpose of hyperfractionating subvolumes of the target. This would primarily affect the interfaces between the target and surrounding healthy tissue \cite{Unkelbach2018a,Bennan2021,Unkelbach2021}. Therefore, for the reference jointly optimized plan in this study, photons were used to achieve a steeper dose gradient at the PTV-brain stem interface using the lateral dose fall-off of photons as opposed to the proton distal edge fall-off.

When introducing LMOs to the OAR, the use of photons is additionally driven by the aim to reduce the high LET dose within the brain stem. Both dirty-dose-based objectives and LETxDose objectives seem similarly adept, where photon dose is used to replace proton dose regions in the target that contribute to high LET dose within the brain stem. As the photons are utilized in small volumes at the end of range of the particle therapy beams, conceivably, these plans may be especially suited for delivery with stereotactic and radiosurgery modalities. The photon contribution can be controlled through the manipulation of the objectives strength/ prescribed dose for the LMO. 

The contribution of photon dose to the cumulative dose can be controlled using the penalty of the LET modifying objectives.
By using dirty-dose objectives, the proton contribution decreases while the photon contribution increases, especially around the OAR and in the body. This leads to sparing the healty tissue and keeping the target coverage. For this purpose high LET dose is replaced by low LET dose coming from photons. The strength of the objectives is mainly determined by the penalty: The stronger the penalty, the stronger the influence of the objective and the more the LET is pushed to the edge of the target. The greater the penalty, the smaller the reduction in the dirty-dose. The strength of the objective is not easy to quantify, let alone convert into comparable penalty. The penalty strength causes uncertainty in the assessment of the objectives used and is due to the standardization that enables a clear comparison of the effectiveness of the objectives that are used but is not done. That is why the strength can only be predicted approximately. 

Dirty-dose is mainly defined through the LET threshold. A lower LET threshold has the effect that almost the entire dose is assigned to the dirty-dose. This would result in a much more aggressive objective that would attempt to shift the entire dose distribution. For LET thresholds that are too high, the tangible influence of the dirty-dose objectives would decrease, as hardly any dose would be defined as a dirty-dose. The choice of the LET threshold is therefore decisive and strongly influences the dose and the $LET_d$ distribution. Note that there is an approximation error in the calculation of the dirty-dose. In principle, the LET within a voxel is not a fixed value, as the protons can have different energies. Therefore, the LET should correctly be regarded as a spectrum. However, no spectrum is used in matRad, but an average LET value. This reduces the accuracy of the calculation but since there is no fragmentation tail for protons, the approximation error can be considered small.

The photon contribution of the joint optimization is theoretically also linked with the choice of beam angles, primarily the proton beam angles due to the higher LET contribution compared to the photons. Beam angles whose distal edge of beam is not located in the direction of the OAR  are less influenced by LET modifying objectives. 

A strong effect compared to the dirty-dose objectives is recognizable for the LETxDose objectives, which can also be caused by the prescribed LETxDose, which was set to zero.
If the penalty is increased, the dirty-dose objective reacts by reducing the LET and thus also a reduction in the dirty-dose, while the LETxDose objective increase the dirty-dose and the LET in the PTV to achieve a sharp reduction of the LET in the OAR.

LMOs applied to the Target (GTV) overcome a different challenge. Firstly, the only proton plan and the reference jointly optimized plan both show a large variation in the dirty-dose applied in the GTV as seen in Figure \ref{fig:allSlice} (c) and (g). Comparatively, purely biological optimization with RBE does not affect the high-LET dose contributions in the OARs as strongly.

On the other hand, the jointly optimized plans with LMOs achieve a fairly uniform dirty-dose value in the GTV.  This suggests that the LMOs seem quite useful in guaranteeing a constant base dirty-dose level in the target. This would theoretically reduce the probability of a local treatment failure by ensuring higher LET dose in the target that would not be achievable with any of the RBE models that have been proposed for proton treatments. However, LETxDose based objectives would have limited applicability in the target volume, as it is difficult to identify a sensible prescription as it opens up the option of the optimizer applying greater dose of low LET protons to satisfy objectives. 

The quadratic property of the LETxDose objective and the trade-off between LET shift and dose shift show a broad spectrum of setting options that can be realized based on the penalty and the prescribed LET/dirty dose. Larger and more varied examinations, as well as objective standardization, can better quantify the created objectives. Jointly optimizing combined treatments gives us the opportunity to investigate other tradeoffs in LET without compromising target coverage.
   
This paper does not try to attempt a one-to-one mathematical comparison but rather to give the reader an introduction to LET modifying objectives and their behavior for combined treatments. These objectives are motivated by different rationales and are prescribed with very different ideas in mind. For example, the prescription of 90 $Gy keV/\micro m$ for the LETxDose underdosing objective for the GTV was based on delivering half the prescribed target dose (30~Gy) with 3 $keV/\micro m $ which is approximately the LET at the Bragg peak. Where as the dirty-dose objective guarantees a minimum of 2 $keV/\micro m$ in the GTV. This suggests that the dirty-dose or a similar concept may be a pragmatic approach to LET-based optimizations focusing on high LET.

Currently, the LET threshold (2 $keV/\micro m$) used to determine dirty-dose, is chosen as a safe option within previously published LET range. The LET threshold could be better biologically motivated with in-vitro  and clinical studies. For example, a retrospective study correlating target adjacent toxicity to LET. This might even suggest that different treatment sites might warrant distinct LET threshold values

A limitation of our implementation is the simplified assumption on calculating dirty-dose. The LET within a voxel of the individual beamlet contributions was treated as a single value for dirty-dose thresholding. Ideally, dirty-dose should be obtained by thresholding on the full LET-spectrum of all particles within that voxel for each beamlet. However, for the purposes of optimization, our assumptions allowed for a faster approximation of LET, as performance-optimized Monte Carlo codes for beamlet-based planning including the capabilities for dirty-dose thresholding during scoring were not available at the time of the study.

The concept of a "quasi" $LET_d$ is an artificial quantity that does not relate to a physical meaning ( Equation \ref{eq:Gesamt-LET}). It does, however, allow us to consider multiple heterogeneous fractions of varying LETs and accumulate and estimate a dose averaged mean over multiple fractions. 

Future exploration, should also investigate the implications this method for other particle treatment options (Helium and carbon ions). Unlike proton therapy, in the case of heavier ions used in radiotherapy, the variation in LET, and thus RBE, is much wider for larger portions of the beam. Furthermore, heavier ions have sharper Bragg peaks making it possible to deliver more conformal plans. Therefore the overdosing LMO could show a more limited use, but would be pertinent in cases of complex or "problematic" patient geometry or beam angle selection, thus improving the safety of particle treatments. On the other hand, such objectives show a lot of potential in boosting and homogenizing the LET/ dirty dose within the target, similar to multi ion treatments \cite{Kopp2020}.

Considering the similarities between jointly optimized plans with LMO in the OAR and Target, photon dose is used in regions that would lack dose from the proton fractions. The overdosing objectives in the OAR suppress the distal proton dose and the underdosing objectives in target boost the proximal proton dose. Therefore conceivable the two objectives can be used concurrently and in conjunction to shape LET in the patient. 

Combined treatments with heterogeneous dose distributions as shown here are naturally susceptible to deterioration in plan quality from range and setup uncertainties. For sites like the brain, like the Anaplastic astrocytoma case shown here, this is less of a concern. However when considering other sites and indications, robust optimization strategies must be investigated \cite{Fabiano2020a}.

\section{Conclusion}
This study demonstrates the power of implementing LET-based objectives within a jointly optimized combined proton-photon treatment framework, highlighting the distinct roles of each modality in optimizing dose distribution and reducing the risk for the OAR. Photons are utilized within the optimization process to aid in creating sharper dose gradients and reducing dose regions coming from high-LET, especially in sensitive areas like the brain stem. However, the applicability of these objectives is influenced by various factors such as beam angles and the specific treatment area. The potential for improving target coverage and minimizing toxicity through the LET-optimization in a proton-photon joint optimized plan offers novel treatment options that maybe safer and more effective.

\section*{Acknowledgments}
This work was, in part, funded by the Deutsche Forschungsgemeinschaft
(DFG, German Research Foundation) – Project No.~443188743.

\printbibliography

\end{document}